# Dose Validation of GRID Block Treatment Applicator within the RayStation Treatment Planning System

**Blessing Akah**[1,2], Edwin Quashie[3], Gene Cardarelli[4]

[1] Physics Department, University of Rhode Island, [2] Mary Bird Perkins Cancer Center, [3] Indiana University Medical School, [4] Warren Alpert Medical School of Brown University

## Abstract

Spatially Fractionated Radiation Therapy (SFRT) or GRID therapy has been in existence for more than a century. GRID therapy was invented as an approach to address the challenges posed by bulky tumors. To treat patients with GRID therapy, the protocol needs to be implemented within the Treatment Planning System (TPS) software. Also, the dose needed for treatment must be validated. A few TPS software vendors such as Elekta and Varian have protocols for treating patients with GRID therapy. However, we are the first to implement the protocols within RayStation TPS. To achieve these, we used the .decimal GRID block applicator to create conformal treatment plans. For the beam modeling, several dosimetry data, and measurements such as percentage depth dose (PDD), beam profile, output, and GRID factors, were obtained for different field sizes and energies. The dose distribution from the GRID openings covered the Planning Target Volume (PTV), with the QA plan having a 98% agreement with the planned dose. Also, we used the Mapcheck2 results to verify the adjustments made to the script variables such as aperture size (hole and spacing). We developed a robust way to validate the dose needed for patient treatment using the GRID block applicator. Also, we have proposed a novel approach to standardize the clinical implementation of the GRID block in RayStation TPS.

## Introduction

Tumor management (control) in patients with bulky tumors presents challenges, especially when conventional therapies like external beam radiation therapy (EBRT) face limitations due to tumor size. Spatially fractionated radiation therapy (GRID therapy) emerged as a novel approach to address the challenges posed by bulky tumors. This technique, initially introduced in the early 90's in Germany by Kohler (1874–1947) and then in the United States in 1933, involves delivering alternating high and low-dose rates of radiation to create a nonuniform distribution within the tumor (Laissue, 2012). GRID therapy was commonly used during the period of orthovoltage radiation to reduce surface-level radiation effects (Hams, 1952). However, this practice became outdated with the introduction of megavoltage technology. Nevertheless, the advent of high-dose-rate (HDR) brachytherapy applicators has highlighted the radiobiological significance of GRID therapy: allows the safe administration of higher radiation doses without surpassing the tolerance of healthy tissues.

The radiobiological basis of GRID therapy's success lies in its ability to induce both direct physical and indirect biological effects on tumor cells (Sathishkumar et al. 2002). The most pronounced indirect effect of radiation is the concept of the bystander effect, where irradiated cells induce damage to neighboring non-irradiated cells (Goodwin etal., 1998, Prise etal., 1998, Djordjevic etal., 2000, Belyakov etal., 2001) - plays a significant role in GRID therapy. GRID therapy has demonstrated notable clinical outcomes in treating bulky tumors. The therapy's unique approach, combining radiobiological insights and innovative delivery techniques, has led to a complete response rate of 62.5% and a pathological response rate of 50% in patients (Mohiuddin et al. 1999). Its ability to effectively manage massive and bulky tumors without significant acute morbidity or late effects on skin, gastrointestinal, central nervous system, and subcutaneous tissues is particularly promising (Mohiuddin et al. 1999). Combining GRID therapy with conventional treatments, either before or after, has shown enhanced tumor control compared to standalone or other multi-modal (Fu etal., 2000; Mendenhall etal., 1986/1998; Brizel et al., 1998) approaches.

GRID therapy has been used to treat various tumors, including those in the head and neck, esophagus, pelvis, and lung (Mark etal., 1950 & 1952). The GRID technique has shown advantages in terms of tumor control and sparing of normal tissue damage, particularly when compared to conventional open-field EBRT. Despite using hyper-fractionation to enhance tumor control, it remains inadequate for late-stage and bulky tumors (> 8cm) (Mohammed Mohiuddin et al. 1999). Employing a grid block with 256 holes in a 16 x 16 cm field yielded 50-50% open and closed areas (Mohammed et al. 1999). Measurements of relative and absolute dose for a 6MV beam revealed dose under the grid's blocked area to be 25-30% of that at the center of the grid holes. Total doses range from 10 Gy to 15 Gy, treating tumor sizes of 6 x 5 cm to 25 x 25 cm has been reported, with exceptional tolerances (neither acute side effects nor long-term complications) - unattainable with conventional EBRT due to larger field sizes (Mohammed Mohiuddin et al. 1990).

By using the GRID block, the open radiation field is converted into a series of pencil beams (Majid Mohiuddin and Park 2000), delivering alternating high and low-dose rates of radiation to create a non-uniform, high-dose, pencil beam radiation through a collimated grid with a 50:50 open-to-closed blocked area ratio [1]. By adjusting the radius of the holes, we can get the grid output and penumbra width "right".
By these, we can improve tumor control while reducing toxicity to healthy tissues (OARs). GRID therapy treatment is available when using the Eclipse and Monaco treatment planning system (TPS), but not in the RayStation TPS software. So, in this project, we validated the dose required when using the GRID applicator for patients', with RayStation TPS. treatment. Also, we proposed a novel treatment protocol for standardizing clinical implementation of GRID block for all users utilizing RayStation for their GRID therapy treatment planning software.

## Methods

The GRID block applicator (Fig. 1) was custom-made with brass material from .decimal [1]. The grid block has a total of 149 holes, which comprises 11 holes by seven rows and 12 holes by six rows. The dimensions of the hole from the isocenter are 25cm x 25cm while the total GRID dimension is 30cm x 30cm. By adjusting the radius of the holes, we can get the grid output and penumbra width "right". A script was created in RayStation Treatment Planning System (TPS), describing the. decimal block as one opening containing all holes with sub-mm-wide channels in between. The radiation treatment plan files were generated in RayStation TPS using beam parameters such as isocenter, treatment angles, and jaw positions to meet the prescription of the PTV (Fig. 2).

An EBT3 GafChromic film was placed directly on top of 10cm solid water for backscatter with a build-up of 5cm in solid water for each energy Output and GRID factors at both CAX and OAX (lateral shift=3cm) were obtained at 100cm SSD at dmax of 1.5cm, 2.5cm and 3.0cm using a Markus ion chamber for 6MV, 10MV and 15MV respectively (Fig 3). These absolute dose measurements were done at 95cm SSD.The Percentage depth dose (PDD) for different field sizes and energies was obtained (Fig. 4). Beam profiles at different depths: dmax, 5cm, and 10cm were also obtained in in-plane and cross-plane using a 3-D water tank. A Quality Assurance (QA) plan was delivered on MapCheck2 phantom, and the dose distribution was evaluated using criteria: 3%/3mm at a 10% threshold (Fig. 5 and 6). The linear accelerator (LINAC) used was Varian 21_IX.

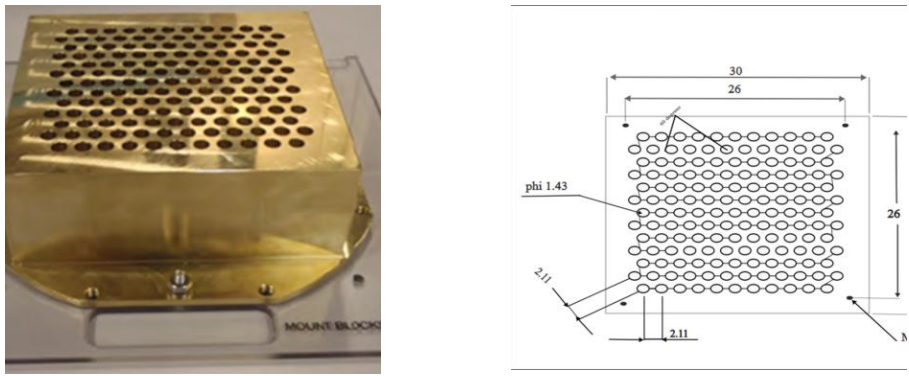

*Fig 1*. Grid block design. (a) The dimensions of the block at the isocenter are 25cm x 25cm. The hole diameter and spacing are 1.43cm and 2.11cm, respectively. (b) Actual Brass GRID block and weighs about 34.2 pounds.

## Results

The dose distribution for a lung test patient was obtained by running the script file in the RayStation treatment planning system (TPS). Dose from the GRID block (openings and blocked area) totally covered the planning target volume (PTV); the hot spot (peak dose) fell within the PTV. Although, small fraction of the cold spot (low dose) fell within the PTV, the higher dose was enough to cause tumor with reduced toxicity to the organ at risk (OARs).

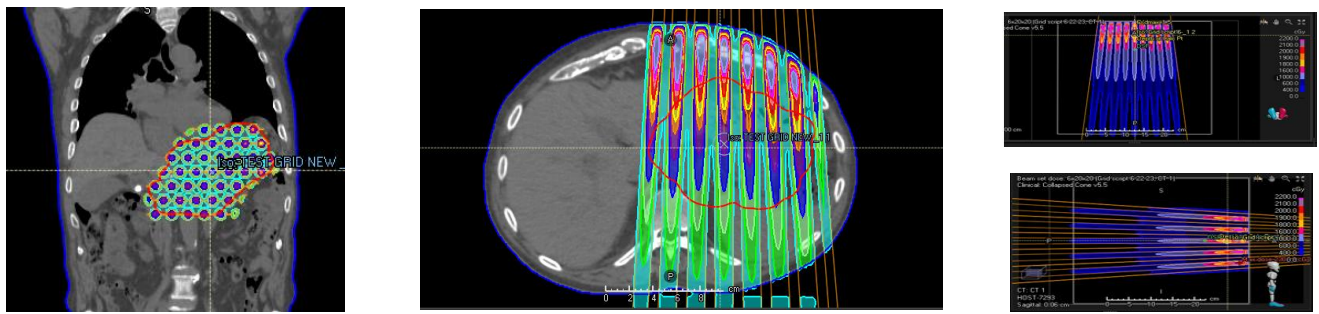

*Fig. 2. (left) Beams Eye View shows the GRID block pattern on a patient's PTV. The GRID openings cover the PTV. (center) Dose distributions showed different isodose line levels with PTV coverage. **Right**: Beam pattern showing GRID openings. (top) AP direction. (bottom) Lateral direction.*

Using EBT3 GafChromic film we were able to measure the absolute dose and obtain the transmission from the GRID block (blocked area). The transmission from was scaled in the RayStation from 0.01 to 0.025. The horizontal profile showed a more symmetrical dose profile than the vertical profile due to the beam projection from the block. The penumbra of the beam energies was also evaluated using the film results.

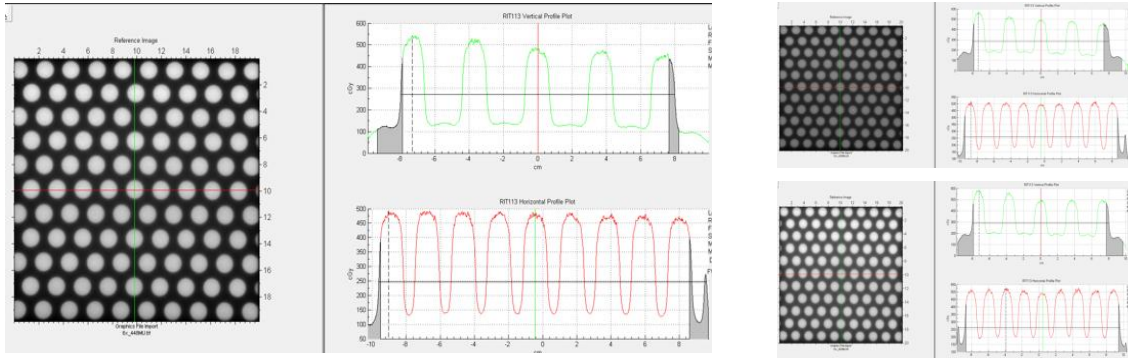

*Fig 3. Film absolute dose results for 6MV (**left-big**), 10MV (**top right**), 15MV (**bottom right**) for 20cm x 20cm GRID at 5cm depth in solid water phantom.*

This section was needed for beam modeling. The Monte Carlo simulation obtained from the TPS is matched with the actual measurements from the linear accelerator (LINAC). Output factor without GRID (open field) and GRID factor with the GRID block were consistent: increases with an increase in the field size for all energies (6, 10 and 15 MV). This increase is due to more scattering at larger field sizes. The percentage depth dose (PDD) value increases with increased energy and depth of maximum dose (dmax) extends deeper as the energy increases. The beam profile showed peak and valley dose corresponding to the open and blocked area of the GRID respectively. The measured dose decreases with depth, with depth of dmax (1.5 cm for 6MV energy) having the highest dose while the depth of 10cm has lowest peak.

### Table 1

|  | CAX | | | OAX | | |
|---|---|---|---|---|---|---|
| FS | 6MV | 10MV | 15MV | 6MV | 10MV | 15MV |
| 5 x 5 | 0.9287 | 0.9424 | 0.9391 | 0.0678 | 0.1356 | 0.1520 |
| 10 x 10 | 1.0000 | 1.0000 | 1.0000 | 1.0000 | 1.0000 | 1.0000 |
| 15 x 15 | 1.0428 | 1.0292 | 1.0324 | 1.0575 | 1.0373 | 1.0394 |
| 20 x 20 | 1.0855 | 1.0480 | 1.0533 | 1.0965 | 1.0573 | 1.0617 |
| 25 x 25 | 1.1120 | 1.0600 | 1.0671 | 1.1273 | 1.0704 | 1.0758 |

### Table 2

|  | CAX | | | OAX | | |
|---|---|---|---|---|---|---|
| FS | 6MV | 10MV | 15MV | 6MV | 10MV | 15MV |
| 5 x 5 | 0.9408 | 0.8477 | 0.8281 | 0.8788 | 0.7643 | 0.7585 |
| 10 x 10 | 0.9043 | 0.8233 | 0.8054 | 0.2752 | 0.3825 | 0.3977 |
| 15 x 15 | 0.8867 | 0.8212 | 0.8026 | 0.2893 | 0.3932 | 0.4081 |
| 20 x 20 | 0.8724 | 0.8222 | 0.8031 | 0.2996 | 0.4012 | 0.4159 |
| 25 x 25 | 0.8700 | 0.8281 | 0.8082 | 0.3060 | 0.4106 | 0.4250 |

**Table 1**: shows the output factors measurement for CAX and OAX without GRID.
**Table 2**: shows GRID output factors for CAX and OAX for the 3 energies.

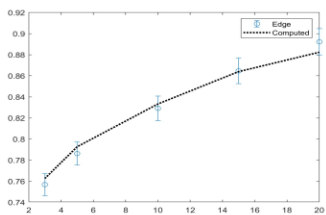
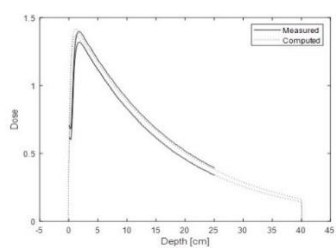
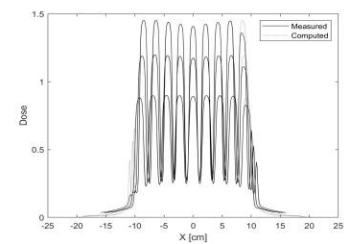

*Fig. 4. 6 MV energy (a) Output factor at 10 cm depth for different field sizes (3x3 cm to 20x20 cm). (b) The PDD for 10x10 cm (lower curve) and 20x20 cm (higher curve). (c) The beam profile at different depths: dmax (highest peak), 5cm (middle peak), and 10cm (lowest peak).*

We validated the treatment plan dose to ensure accuracy during delivery. The criteria 3%/3mm at 10% threshold showed a 98% passing rate gamma analysis.

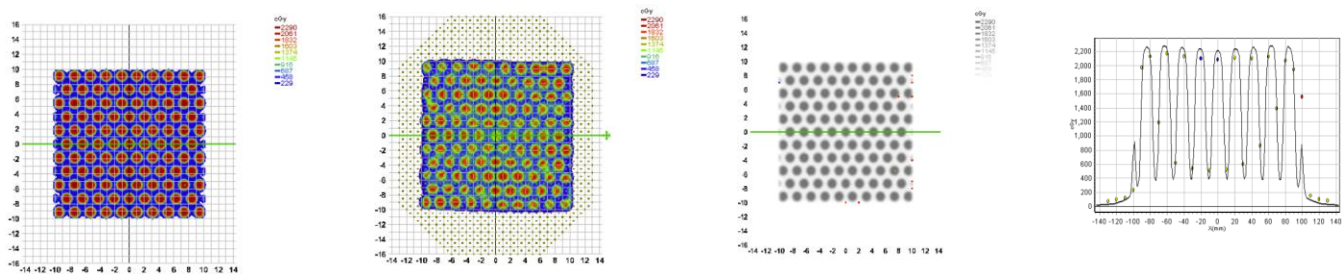

*Fig 5. Mapcheck2 results for 6 MV energy (a) Planned dose. (b) Measured dose. (c and d) gamma passing rate of 98.0% at 3%/3mm at a 10% threshold for 20cm x 20cm.*

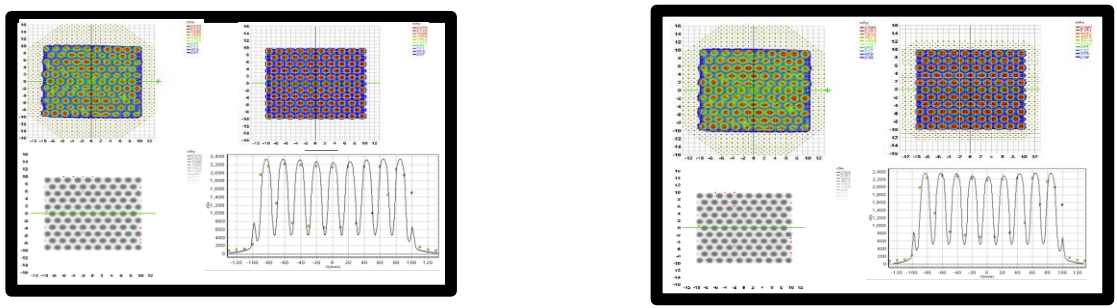

*Fig 6: Mapcheck2 results for 20x20 cm field size: 10 MV energy (**left**) and 15 MV energy (**right**)*

## Discussion

In this clinical project, we developed a robust way to validate the dose needed for a patient treatment by considering the output and GRID block factors. This project was performed at the Rhode Island Hospital (RIH) with all the measurements and data obtained using Varian TrueBeam. The script and beam modeling part of this project was done by the senior physicist at RaySearch Laboratory to create a GRID treatment plan. After the script was completed, we ran the script and achieved total coverage on the planning target volume (PTV).

Next, we obtained percentage depth dose (PDD), beam profile and output factor (with and without GRID block). All these dosimetry measurements were supplied to accurately model the beam parameters within the RayStation treatment planning system (TPS). This data was collected multiple times to get close results with the simulated results from the TPS. To account for the small penumbra of the beam, we utilized edge detector due to its small detector area compared to the farmer's ionization chamber. The absolute dose was also analyzed and measured using the GafChromic film, which allowed the transmission from the grid block to be scaled up from 0.01 to 0.025. The results obtained from these measurements were like simulated data from the RayStation. After this, the planned dose was validated on Mapcheck 2 using criteria 3%/3mm at 10% threshold. The gamma passing rate was 98%, which means that the LINAC was 98% accuracy in delivery the planned dose. Next will be to validate the dose on a "real" patient, after which we can start treating eligible patients with GRID block applicator at the Rhode Island Hospital.

Also, we have proposed a novel approach to standardize the clinical implementation of the GRID block in RayStation in an easier and accurate way to treat bulky tumors as a single treatment or as a boost, without sacrificing dose to OARs. Other clinics across the United States and even abroad can safely utilize our protocol to implement GRID therapy. The RaySearch Laboratory will be including this protocol in their upcoming TPS software release later this year, 2024. Our future development is to use this validation process to script an MLC solution to replace the GRID block. Although this method has an extended treatment time and higher valley-to-peak ratio, it would be much more convenient to use and cost efficient.

## Conclusions

GRID therapy presents a groundbreaking method for tackling the challenges associated with treating bulky tumors. We have developed a robust way to validate the dose needed for a patient treatment by considering the output and GRID block factors. Also, we have proposed a novel approach to standardize the clinical implementation of the GRID block in RayStation in a safer and more accurate way to treat bulky tumors as a single treatment or as a boost, without sacrificing dose to OARs. Our future development is to use this validation process to script an MLC solution to replace the GRID block.